\input{aipcheck}

\documentclass[
    ,final            
  ]
  {aipproc}

\layoutstyle{6x9}

\begin{document}

\title{Latest Results on $g_1$ and $g_2$ at high $x$}


\author{Jian-ping Chen}
{
  address={Jefferson Lab, Newport News, Virginia 23606, USA}
}


\begin{abstract}
Recent progress from Jefferson Lab has significantly improved our 
understanding of the nucleon spin structure in the high-$x$ region. 
Results of a precision measurement of the neutron spin 
asymmetry, $A_1^n$, in the high-$x$ (valence quark) region are discussed. 
The 
up and down quark spin distributions in the nucleon were extracted. 
$A_2^n$ was also measured. The results
were used, in combination with existing data, 
to extract the second moment, $d_2^n$.
Preliminary results on $A_1^p$ and $A_1^d$ in the high-$x$
region have also become available. 
Finally, the results of a precision measurement of 
the $g_2$ structure function to study higher twist effects will be presented. 

\end{abstract}

\maketitle


\subsection{Introduction and Motivation}

Recently, the high polarized luminosity available at Jefferson 
Lab (JLab) has allowed the study of the nucleon spin structure with 
an unprecedented precision, enabling us to access the
valence quark (high-$x$) region and also to expand the study to the
second spin structure function, $g_2$.

The high-$x$ region is of special interest, because this is where the valence 
quark contributions are expected to dominate.
With sea quark and explicit gluon contributions expected not to be
important, it is a clean region to test our understanding of nucleon
structure. Relativistic constituent quark models~\cite{vqm}
should be applicable in this region
and perturbative QCD~\cite{pQCD} can be used to make predictions in the large
$x$ limit. 

To first approximation, the constituent quarks in the nucleon are
described by the SU(6) wavefunctions.
SU(6) symmetry leads to the following predictions: 

\begin{equation}
A_1^p=5/9;\ \ A_1^n=0; \ \ \Delta u/u=2/3; \ \ 
\Delta d/d=-1/3.
\label{eq:SU6}
\end{equation}

Relativistic Constituent Quark Models (RCQM) with broken SU(6) symmetry, e.g., 
the hyperfine 
interaction model~\cite{vqm}, lead to a dominance of a `diquark' 
configuration 
with the diquark spin $S=0$ at high $x$. This implies that as $x\rightarrow1$:
\begin{equation}
 A_1^p\rightarrow 1;\ \
   A_1^n\rightarrow 1;\ \ \Delta u/u \rightarrow 1;\ \ {\rm and} \ \ 
\Delta d/d \rightarrow -1/3.
\label{eq:rnpqcd}
\end{equation}
\noindent In the RCQM, relativistic effects give rise to 
the quark orbital angular momentum and reduce the valence quark 
contributions to the nucleon spin from 1 to $0.6 - 0.75$.
 
Another approach is leading-order pQCD~\cite{pQCD}, which assumes the 
quark orbital angular momentum to be negligible and leads to hadron helicity 
conservation. 
It yields:  

\begin{equation}
A_1^p\rightarrow 1;\ \
   A_1^n\rightarrow 1;\ \ 
\Delta u/u \rightarrow 1;\ \ {\rm and} \ \ 
\Delta d/d \rightarrow 1.
\label{eq:rnppqcd}
\end{equation}
\noindent
Not only are the limiting values as $x\rightarrow 1$ important, but also
the behavior in the high-$x$ region. How $A_1^n$ and  $A_1^p$ 
approach their limiting values when $x$ approaches 1, is sensitive to
the dynamics in the valence quark region. 

$g_2$, unlike $g_1$ and $F_1$, cannot be
interpreted in the simple quark-parton model. To understand $g_2$ properly, it is best to 
start with the operator product expansion 
method (OPE)~\cite{OPE}.
In the OPE, neglecting quark masses, $g_2$ can be cleanly separated into a
twist-2 and a higher twist term:
  \begin{eqnarray}g_2(x,Q^2) = g_2^{WW}(x,Q^2) +g_2^{H.T.}(x,Q^2)~.
  \end{eqnarray}
The leading-twist term can be determined from 
$g_1$ as~\cite{WW}
  \begin{eqnarray}
   g_2^{WW}(x,Q^2) = -g_1(x,Q^2) + \int _{x}^1 \frac{g_1(y,Q^2)}{y} dy~,
  \end{eqnarray}
and the higher-twist term arises from the quark-gluon
correlations.
Therefore $g_2$ provides a clean way to study higher-twist effects.
In addition, at high $Q^2$, the $x^2$-weighted moment, $d_2$, 
is a twist-3 matrix element and is related to the color 
polarizabilities~\cite{d2}:
\begin{equation}
d_2 = \int _{0}^{1} x^2 [g_2(x)-g_2^{WW}(x)] dx.
\end{equation}
Predictions for $d_2$ exist from various models and lattice QCD.

\subsection{Recent results from Jefferson Lab}


In 2001, JLab experiment E99-117~\cite{e99117} was carried out in Hall A
to measure $A_1^n$ with high precision in the $x$ region from 0.33 to 0.61
($Q^2$ from 2.7 to 4.8 GeV$^2$). 
Asymmetries from inclusive scattering of 
a highly polarized 5.7 GeV electron beam 
on a high pressure ($>10$ atm) (both longitudinally and
transversely) polarized $^3$He target were measured. 
Parallel and perpendicular asymmetries
were extracted for $^3$He. After taking into account the beam and target 
polarization and the dilution factor,
they were combined to form $A_1^{^3He}$. Using the most recent 
model~\cite{model}, nuclear
corrections were applied to extract $A_1^n$. The results on $A_1^n$
are shown in the left panel of Fig. 1. 

The experiment greatly improved the precision
of data in the high-$x$ region, providing the first evidence that 
$A_1^n$ becomes positive at large $x$, showing clear SU(6) symmetry 
breaking. The results are in good agreement with the LSS 2001 pQCD
fit to previous world data~\cite{LSS2001} (solid curve) and 
the statistical model~\cite{stat} (long-dashed curve).
The trend of the data is consistent with the RCQM predictions
(the shaded band). The data disagree with the predictions from the 
leading-order pQCD models (short-dashed and dash-dotted curves).

\begin{figure}[!ht]
\parbox[t]{0.5\textwidth}{\centering\includegraphics[bb=-20 -28 402 455, angle=0,width=0.5\textwidth]{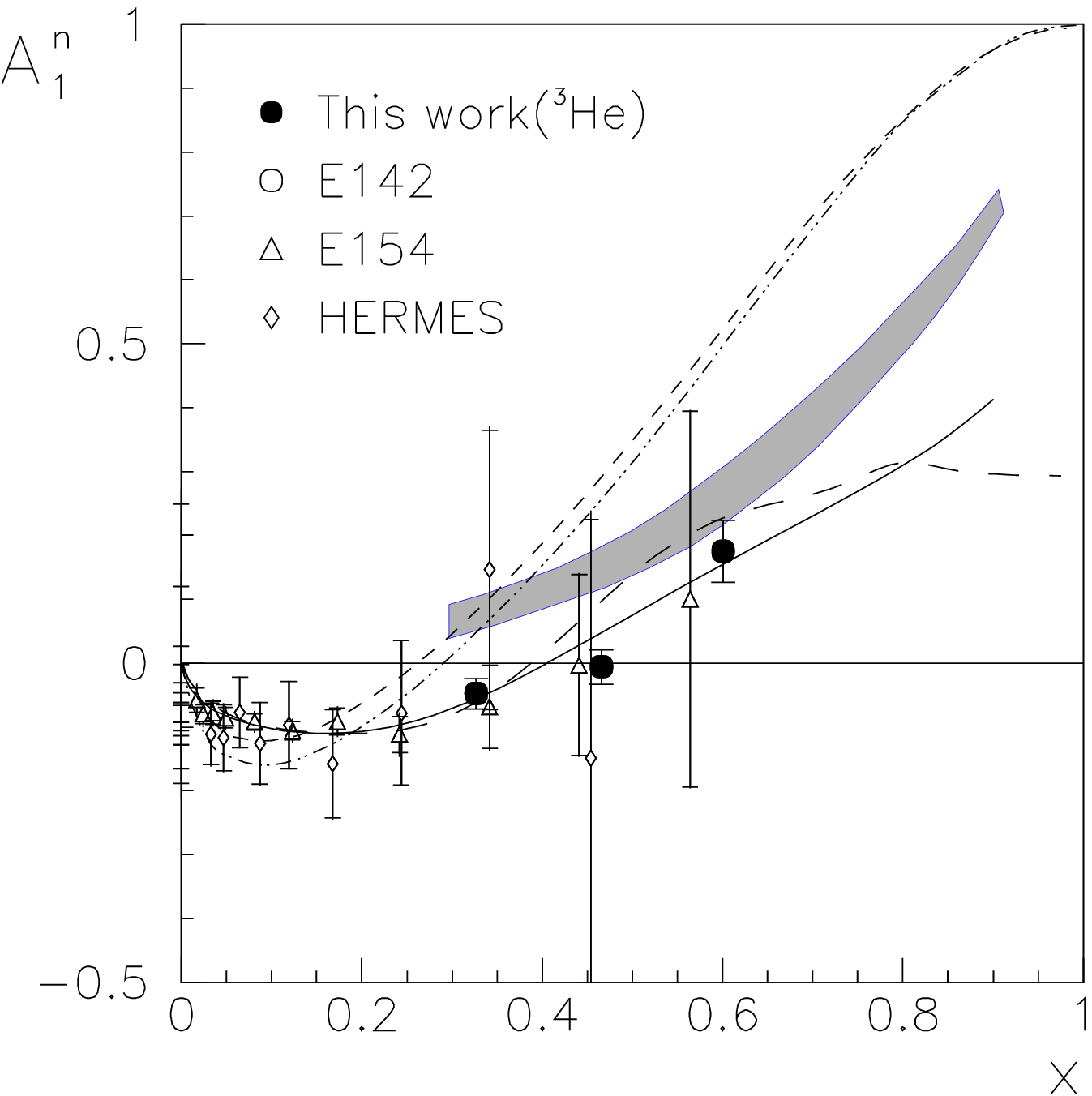}}
\parbox[t]{0.5\textwidth}{\centering\includegraphics[bb=-70 -28 352 455, angle=0,width=0.5\textwidth]{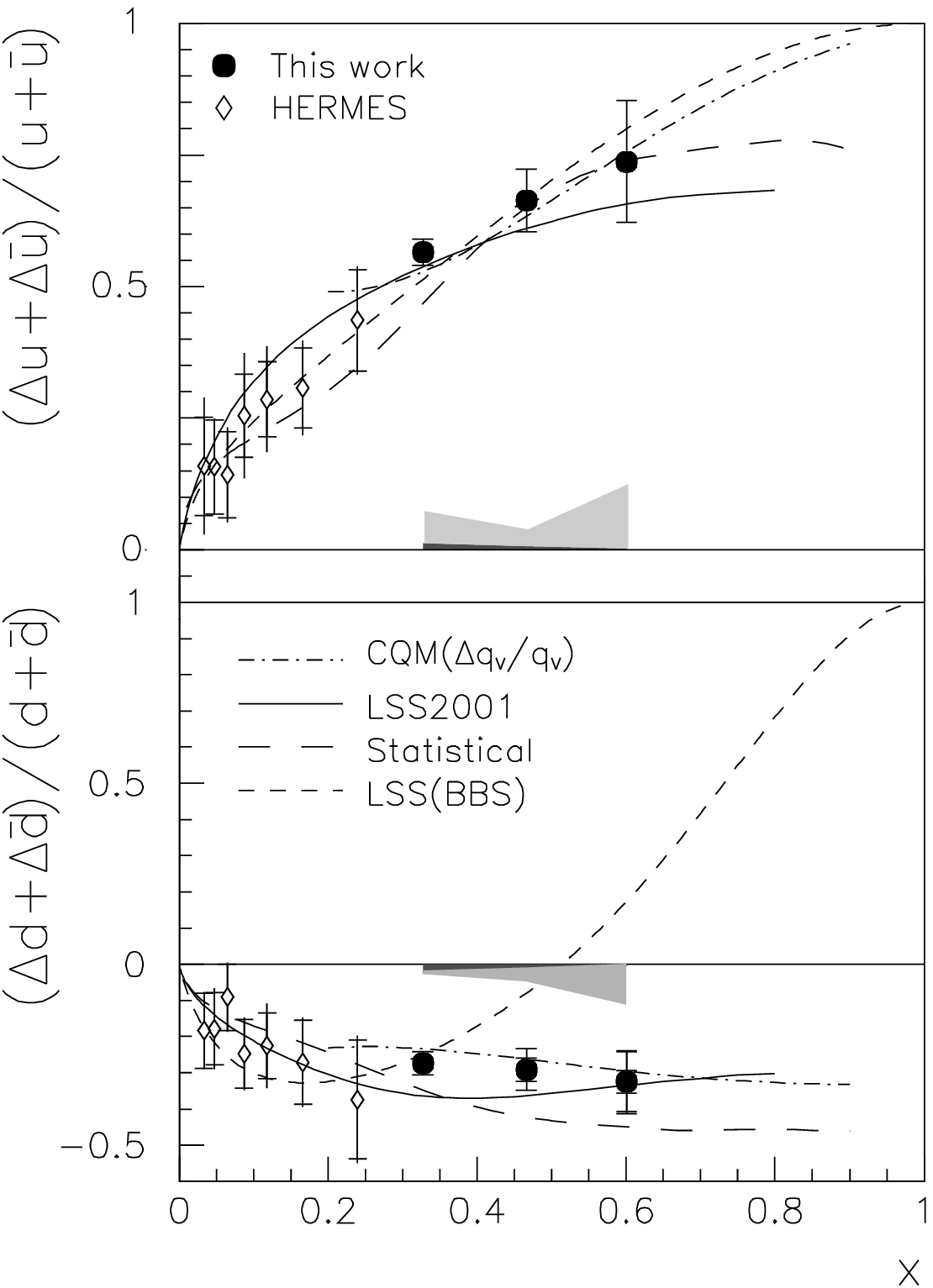}}
\caption  {A$_1^n$, $\Delta u/u$ and $\Delta d/d$ results compared with 
the world data and theoretical predictions.}
\end{figure}
\medskip

In the leading-order approximation,
the polarized quark distribution functions $\Delta u/u$ and $\Delta d/d$ were 
extracted from our neutron data combined with the world proton data. 
The results are shown
in the right panel of Fig. 1, along with predictions from the RCQM (dot-dashed
curves), leading-order pQCD (short-dashed curves), the LSS 2001 fits 
(solid curves) and the statistical model (long-dashed curves).  The results 
agree well with RCQM predictions as well as the LSS 2001 fits and statistical 
model but are in significant disagreement with the 
predictions from
the leading-order pQCD models assuming hadron helicity conservation. This
suggests that effects beyond leading-order pQCD, such as the quark orbital
angular momentum, may play an important role in this kinematic region.

$A_2^n$ was also obtained from the same experiment. The precision of the $A_2^n$ data is comparable to that of the 
best existing world data~\cite{E155x} at high x. Combining these results with the world data, the second moment $d_2^n$ was extracted at an average $Q^2$ of 5
GeV$^2$:

\begin{equation}
d_2^n = 0.0062 \pm 0.0028.
\label{eq:d2nres}
\end{equation}
Compared to the previously published result~\cite{E155x}, the uncertainty on $d_2^n$ has 
been improved by about a factor of 2. The $d_2$ moment at high $Q^2$ 
has been calculated by Lattice QCD and a number
of theoretical models. While a negative or near-zero value was 
predicted by Lattice QCD and most models, the new result for $d_2^n$ 
is positive.

Preliminary results of $A_1^p$ and $A_1^d$ from the Hall B eg1 experiment~\cite{eg1} 
have recently become available. The data cover the $Q^2$ range of 1.4 to 4.5 
GeV$^2$ for $x$ 
from 0.2 to 0.6 with an invariant mass larger than 2 GeV. The precision of the data
improved significantly over that of the existing world data. 


A precision measurement of g$_2^n$ from JLab E97-103~\cite{e97103} covered 
five different Q$^2$ values from 0.58 to 1.36 GeV$^2$ at x $\approx 0.2$. 
Results for $g_2^n$ as well as $g_1^n$ are given in 
Fig. 2. The light-shaded area in the two plots 
gives the leading-twist contribution
to these two quantities, respectively, obtained by fitting world data and
evolving to the $Q^2$ values of this experiment. The systematic errors are 
shown as the dark-shaded area near the horizontal axes.   

\noindent
\begin{figure}[!ht]
\parbox[t]{0.3\textwidth}{\centering\includegraphics[bb=130 322 582 725, angle=-90,width=0.3\textwidth]{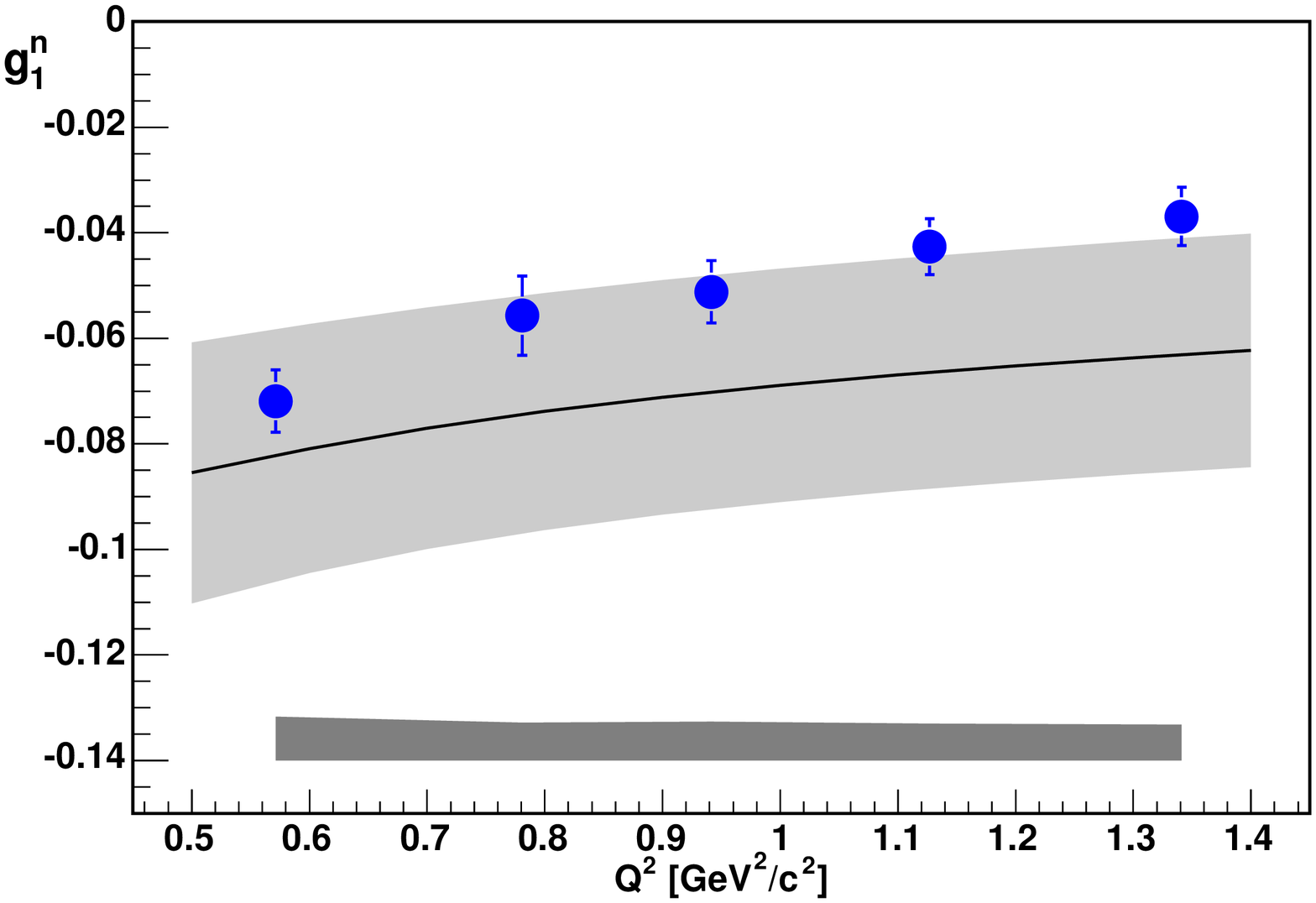}}
\parbox[t]{0.3\textwidth}{\centering\includegraphics[bb=130 50 582 435, angle=-90,width=0.3\textwidth]{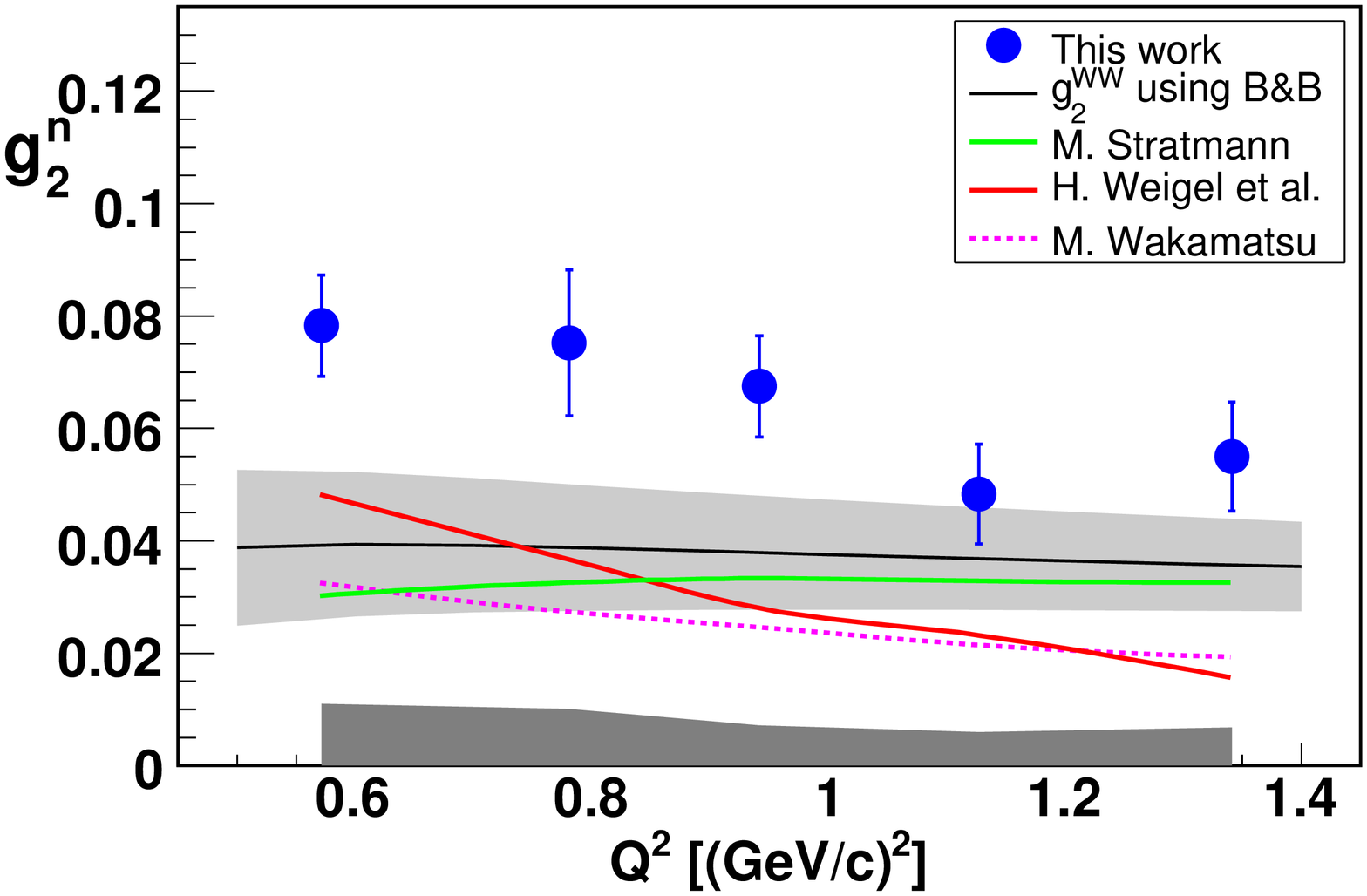}}
\caption{
        Fig.~1: results for $g_1^n$ (left) and $g_2^n$ (right)
from E97103.}
\end{figure}

The precision reached is more than an order
of magnitude improvement over that of the best world data.  The difference 
of g$_2$ from the leading twist part (g$_2^{WW}$)\cite{WW} is due to 
higher twist effects and is sensitive to quark-gluon correlations. 
The g$_2^{WW}$ values were obtained from a fit~\cite{BB} to the world high $Q^2$ data,
then evolved to the $Q^2$ values of this experiment. 
The measured g$_2^n$ values 
are consistently higher than g$_2^{WW}$.
For the first time, there is a clear indication that higher twist effects 
become important at the level of precision of these data. 
The new $g_1^n$ data agree with  
the leading-twist calculations within the uncertainties.

In summary, the high polarized luminosity available at
JLab, has provided high-precision data to study the nucleon
spin structure in the high-$x$ region and higher twist effects, which shed
light on the valence quark structure and help to
understand quark-gluon correlations.



\footnotesize{
The work presented was supported in part 
by the U. S. Department of Energy (DOE)
contract DE-AC05-84ER40150 Modification NO. M175,
under which the
Southeastern Universities Research Association operates the 
Thomas Jefferson National Accelerator Facility.
}





\end{document}